\documentclass[twocolumn,showpacs,preprintnumbers,amsmath,amssymb,prb]{revtex4}
\usepackage{amsmath,amsfonts,bm,graphicx}
\usepackage{color}

\newcommand{\VVV}{\mathbf{V}}
\newcommand{\HHH}{\mathbf{H}}
\newcommand{\GGG}[2]{\mathbf{G}^{(#1)}_{#2}}
\newcommand{\GnP}{\mathbf{G}_{n+1}}
\newcommand{\Enm}[2]{\mathbf{E}^{(#1)}_{\mathrm{#2}}}

\begin{document}

\title{Hofstadter butterflies and magnetically induced band gap quenching in graphene antidot lattices}

\author{
Jesper Goor Pedersen$^1$
and Thomas Garm Pedersen$^{2,3}$
}
\affiliation{
$^1$
DTU Nanotech -- Department of Micro- and Nanotechnology, Technical University of Denmark, DK-2800 Kgs. Lyngby, Denmark
\\
$^2$
Department of Physics and Nanotechnology,
Aalborg University, Skjernvej 4A
DK-9220 Aalborg East, Denmark
\\
$^3$
Center for Nanostructured Graphene (CNG), Aalborg University, DK-9220 Aalborg East, Denmark
}
\date{\today}

\begin{abstract}
We study graphene antidot lattices (GALs) in magnetic fields. Using a tight-binding model and a recursive Green's function technique that we extend to deal with periodic structures, we calculate Hofstadter butterflies of GALs. We compare the results to those obtained in a simpler gapped graphene model. A crucial difference emerges in the behaviour of the lowest Landau level, which in a gapped graphene model is independent of magnetic field. In stark contrast to this picture, we find that in GALs the band gap can be completely closed by applying a magnetic field. While our numerical simulations can only be performed on structures much smaller than can be experimentally realized, we find that the critical magnetic field for which the gap closes can be directly related to the ratio between the cyclotron radius and the neck width of the GAL. In this way, we obtain a simple scaling law for extrapolation of our results to more realistically sized structures and find resulting quenching magnetic fields that should be well within reach of experiments.
\end{abstract}


\maketitle

\section{Introduction}

Semiconductor antidot lattices have revealed a range of intriguing transport phenomena. In particular, transport in magnetic fields has been applied in studies of magneto-resistance, localization, Shubnikov - de Haas oscillations, and quantum Hall effects.\cite{PhysRevLett.70.4118, PhysRevB.41.12307} Traditionally, such structures have been based on GaAs heterojunctions. Recently, however, graphene antidot lattices (GALs) have been proposed\cite{ar_tgp1,ar_tgp2} and magneto-transport studies have been realized experimentally.\cite{shen2008, Eroms2009, Shimizu2012, Giebers2012}
Whereas GaAs antidots are typically produced by dry etching or local ion implantation in molecular beam epitaxy grown heterostructures, GALs are produced by simply etching arrays of holes into graphene sheets placed on suitable insulating substrates. Etch masks can be fabricated using either e-beam lithography or block co-polymers. In this manner, features of the order 10--20 nm have been obtained\cite{Kim2010, Bai2010}. In addition to GALs fabricated via etching, the patterned hydrogen adsorption technique\cite{ar_Balog} also produces structures resembling periodic antidots. Here, hydrogenated graphane-like islands form 'forbidden' regions similar to holes but on a few-nanometer scale. Very recently, direct e-beam writing of small GALs with holes around 2 nm has been demonstrated.\cite{Xu2013}

While magneto-transport is well documented in experimental GAL studies, the theoretical description of extended GALs in magnetic fields is a challenging task, because the associated vector potential breaks translational invariance. Structures comprising only a single or few antidots are more easily analysed. For instance, Yang and coworkers\cite{Park2010, Kim2012} have included magnetic fields in studies of isolated graphene antidots and small arrays, with perforations modelled as circular electrostatic potentials. The present authors studied an isolated antidot using a mass term barrier.\cite{Isolated_antidot2012} The problem of modeling periodic arrays of graphene antidots in magnetic fields is much more involved, however. The energy spectrum of electrons in periodic potentials subject to magnetic fields takes the form of self-similar Hofstadter 'butterflies'.\cite{Hofstadter1976} Hofstadter butterflies have previously been studied in pristine graphene,\cite{Faraday_graphite, Hasegawa2006} bilayer graphene,\cite{Nemec2007} twisted bilayer graphene,\cite{Wang2012} graphene with point defects\cite{Islamoglu2012} and graphene quantum dots.\cite{Zhang2008} Very recently, Hofstadter spectra have been studied experimentally in single and bilayer graphene on hexagonal boron nitride.\cite{Dean2012,Ponomarenko2012} In all cases, an intriguing self-similar structure emerges whenever the characteristic magnetic length becomes comparable to the geometrical period. The relatively large size of this period is challenging for an atomistic description of the spectrum, however. In pristine graphene, a magnetic field leads to a gapped Landau level structure. Hence, for graphene modified in order to induce band gaps (such as GALs, electrically biased bilayers,\cite{Castro2007} quantum dots,\cite{Zhang2008} and geometrically sharp electrostatic gates\cite{Zhang2010,Pedersen2012}) an interesting interplay between geometric and $B$-field induced band gaps is expected. A simplified model capturing the essentials of geometrically induced band gaps is that of gapped graphene\cite{Gapped2009} for which a unique Landau level structure\cite{Gusynin2007} and magneto-optical response \cite{Magneto_gapped2011} have been predicted. For magnetic effects, though, little is known about the reliability of gapped graphene as an approximation to graphene with a gapped spectrum induced by geometric modifications.
 
In the present work, we study the properties of GALs in magnetic fields using a tight-binding model with magnetic effects included via a Peierls phase. We compare true antidot geometries to gapped graphene approximations and find important differences. Most importantly, we observe band gap quenching by the magnetic field in GALs whereas in gapped graphene the band gap is completely robust for arbitrarily large fields. Hence, the behavior of GALs is reminiscent of graphene quantum dots\cite{Zhang2008} and graphene nanoribbons.\cite{Huang2007, Kumar2010} We also demonstrate simple scaling laws to extrapolate our results to experimentally feasible structures.
 
\section{Theory and methods}

A GAL is modelled as an infinite, periodic array of circular holes in the graphene sheet. The superlattice spanning the holes is assumed to have regular triangular synnetry, in order to ensure the existence of a full band gap for all unit cell geometries.\cite{ACSnano2011,Ouyang2011}
We denote the radius of the hole and the side-length of the unit cell by $R$ and $L$, respectively, both of which are measured in units of the graphene lattice constant $a$. Hence, a given GAL is designated by $\{L, R\}$, where $L$ (but not necessarily $R$) is an integer.

We use a nearest-neighbor tight-binding model to study the energy spectrum of $\pi$-electrons within the GAL, i.e. the hopping integral $t_{ij}$ is taken as $-t$ with $t=3$~eV for neighboring sites and vanishes otherwise. The effect of the magnetic field is included via a Peierls phase added to the hopping term between atomic sites $i$ and $j$: $t_{ij}\rightarrow t_{ij} e^{i\phi}$, with the phase given as
$\phi=(e/\hbar)  \int_{\mathbf{R}_i}^{\mathbf{R}_j}\mathbf{A}\cdot d\mathbf{l}$. Here, $\mathbf{R}_i$ and
$\mathbf{R}_j$ denote the positions of the atomic sites, while $\mathbf{A}$ is the magnetic vector potential. The graphene sheet is in the $xy$ plane and the magnetic field is taken to be constant and directed perpendicularly to the sheet. Using the Landau gauge $\mathbf{A}=B x \hat{y}$ the Peierls phase becomes

\begin{equation}
\phi=\frac{e B}{2\hbar} (x_i+x_j)(y_j-y_i).\label{eq:peierls}
\end{equation}

\begin{figure}
\begin{center}
\includegraphics[width=\linewidth]{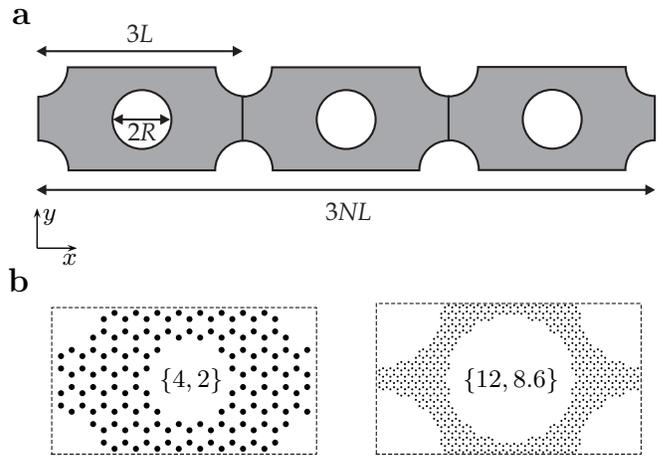}
\caption{(a) Enlarged supercell used for calculating the magnetic properties of GALs. The smallest rectangular GAL unit cell is repeated $N$ times to ensure periodicity of the Peierls phase in the hopping terms of the tight-binding Hamiltonian.
(b) Rectangular unit cells for two examples of GALs with similar neck widths.
}
\label{fig:supercell}
\end{center}
\end{figure}

The crucial advantage of the Peierls phase approach is that lattice periodicity can be restored provided a suitable "magnetic supercell" containing several original unit cells is constructed. To this end, we first construct a rectangular unit cell containing two antidots. This doubled cell is then repeated $N$ times in the $x$ direction as shown in Fig.~\ref{fig:supercell}. Periodicity is subsequently ensured by requiring that the phase shift difference be an integer multiple of $2\pi$ for a pair of neighbor sites at the left- and rightmost ends of the supercell, i.e. separated by a distance of $3 N L a$. The smallest separation along the $y$ axis for coupled atoms is $a/2\sqrt{3}$ and so the minimum $B$-field required for periodicity is

\begin{equation}
B_\mathrm{min}=\frac{2 h}{\sqrt{3} e N L a^2}.\label{eq:Bmin}
\end{equation}

When the flux $\Phi = B\sqrt{3}a^2/2$ through a graphene unit cell equals a flux quantum $\Phi_0=h/e$, the energy spectrum has been restored to the unperturbed one. Writing $B=n B_\mathrm{min}$ with $n$ integer we find the relative flux
$\frac{\Phi}{\Phi_0}=\frac{n}{N L}$.
We can thus calculate the Hofstadter spectra by varying $n$ over the range $\left[0,NL\right]$. In practice, advantage is taken of the fact that the same spectrum is obtained for identical ratios $n/N$, and thus calculations can be performed on much smaller magnetic unit cells whenever $n$ and $N$ are commensurate.

From Eq.~(\ref{eq:Bmin}) it follows that the number of repeated unit cells in the supercell required to handle a certain magnetic field is inversely proportional to the field strength. Because of the relatively large size of the fundamental unit cell of a GAL, the number of atoms in the magnetic supercell becomes very large even for substantial magnetic fields. We thus cannot easily rely on standard diagonalization techniques to determine the properties of GALs in magnetic fields. Instead, we expand on well-known recursive Green's function methods.~\cite{MacKinnon1980,MacKinnon1985} These methods are
commonly applied to transport studies of finite or semi-infinite structures sandwiched between semi-infinite leads. Here, periodicity of the central region is at most along a single direction and the Hamiltonian can easily be written in block-diagonal form, which is the basis upon which the recursive Green's function formalism is built. In our case, periodicity is along both directions. We note that one immediate solution to this problem is to perform a transformation of the Hamiltonian to force it into block-diagonal form, which is possibly even with periodic boundary conditions in both directions. This, however, requires a doubling of the size of the individual cells used in the recursive algorithm which, for large GALs, we find to be slower than our method for calculating the density of states. In the appendix, we present our extension of the recursive Green's function formalism so that periodicity can be handled without reshuffling the elements of the Hamiltonian. We use this formalism to calculate the density of states according to 
$\rho(E) = -\pi^{-1}\mathrm{Im}\{\mathrm{Tr}[\hat{G}^r]\}$, 
where the retarded Green's function 
$\hat{G}^r = (E+i\gamma-\hat{H})^{-1}$ 
includes a small broadening term $\gamma$. A related method based on stochastic evaluation of $\hat{G}^r$ has previously been used to study magnetic effects in twisted bilayers.\cite{Wang2012} 

\section{Hofstadter butterflies}
\begin{figure}
\begin{center}
\includegraphics[width=.95\linewidth]{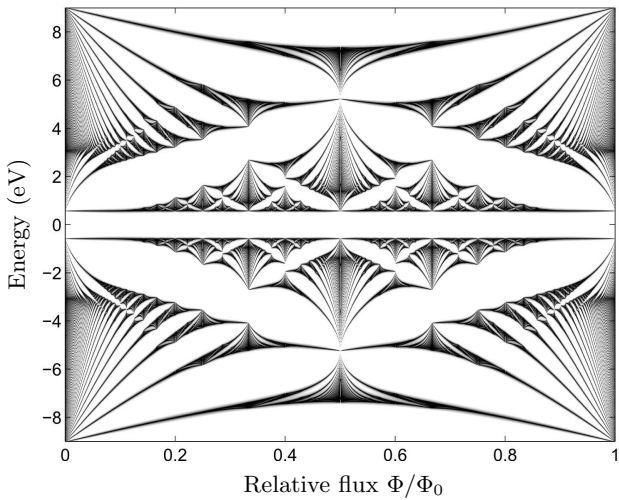}
\caption{(Color online)
Hofstadter butterfly for the gapped graphene model, with the gap set to match that of the $\{4,2\}$ antidot lattice. Shown is the density of states as function of energy and the relative flux through a graphene unit cell. To ease visibility, the logarithm of the density of states is plotted. The bottom panel shows a closer view of the low-flux part of the spectrum. The dashed lines show the lowest five Landau levels expected from a Dirac treatment of the problem.
}
\label{fig:gapped}
\end{center}
\end{figure}

As mentioned above, a phenomenological gapped graphene model may be used as an approximation of the true GAL geometry. Here, alternating on-site energies $\pm\Delta$ are assigned to the atoms belonging to the two sublattices. This model produces a band gap of $E_g=2\Delta$, which is adjusted to the gap of the full GAL band structure. For the optical response, we have previously demonstrated good agreement between the GAL and gapped models for photon energies close to the band gap.\cite{Gapped2009} To compare with the magnetic spectra for GALs we show in Fig.~\ref{fig:gapped} the Hofstadter butterfly of a gapped graphene model, wherein the effect of the antidot lattice is included via such a staggered potential. 
The figure shows the density of states as a function of energy and the relative flux through a graphene unit cell. As expected from the discussion above, we see that when $\Phi/\Phi_0=1$ the spectrum in absence of a magnetic field is restored.
One striking characteristic that emerges in the figure are the zeroth Landau levels, which for gapped graphene sit at $\pm \Delta$, i.e. exactly separated by the band gap.\cite{Gusynin2007} These zeroth Landau levels are characteristic of graphene and exhibit energies that are independent of magnetic field.\cite{Novoselov2005, Zhang2005, Gusynin2005} Whereas for low magnetic fields, where the Dirac model is applicable, the zeroth Landau levels are indeed completely independent of magnetic field, we do see a broadening and eventual splitting of the levels into multiple bands as the magnetic field becomes sufficiently strong. This is equivalent to what is seen for ordinary, gapless graphene.\cite{Rhim2012}

Note that the full range of relative magnetic flux shown in the figure corresponds to unrealistically large magnetic fields up to $79$~kT, well beyond the reach of experiments. A zoom of the low-flux part of the butterfly is shown in the bottom panel of Fig.~\ref{fig:gapped}. For comparison we also show the Landau levels expected from a Dirac equation treatment, $E_{ns}=s\sqrt{\Delta^2+n\hbar^2\omega_c^2}$, with $s=\pm 1$ and $n$ a non-negative integer. Here, $\omega_c=\sqrt{2}v_F/l_B$, with $v_F$ the Fermi velocity of graphene and $l_B=\sqrt{\hbar/(eB)}$ the magnetic length.
Note how the fractal structure only really emerges once the relative flux is of the order of $\Phi/\Phi_0=0.05$, corresponding to a huge magnetic field strength of the order of $B\simeq 4$~kT.
The crucial parameter for seeing signatures of the fractal structure is that the magnetic length should be of the order of the lattice constant of the material. In pristine graphene -- and indeed in any bulk material -- this requires unrealistically large magnetic fields. Thus, the much larger superlattice introduced by the GAL is a way of overcoming this obstacle. A similar explanation applies to the recent experimental studies of effects related to Hofstadter butterflies in single- and bilayer graphene on hexagonal boron nitride.\cite{Dean2012,Ponomarenko2012} Here, a  moir\'{e} superlattice is formed due to the slight mismatch in lattice constant between graphene and boron nitride. In the present work, the fundamental superlattice period is given by the GAL lattice constant $\sqrt{3} L a$. Thus, due to the $B^{-1/2}$ behavior of $l_B$, the $B$-field, at which novel magnetic features become visible, is reduced roughly by a factor of $L^2$.

\begin{figure}
\begin{center}
\includegraphics[width=.95\linewidth]{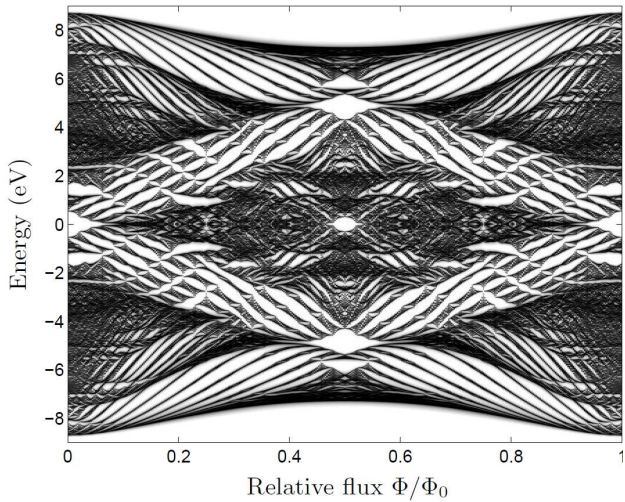}
\caption{
Hofstadter butterfly for the $\{4,2\}$ GAL, showing the density of states as function of energy and the relative flux through a graphene unit cell. To ease visibility, the logarithm of the density of states is plotted. Note the quenching of the gap by the magnetic field around $\Phi/\Phi_0=0.1$, shown in more detail in the zoom in Fig.~\ref{fig:42GALz}.
}
\label{fig:42GAL}
\end{center}

\end{figure}
\begin{figure}
\begin{center}
\includegraphics[width=.95\linewidth]{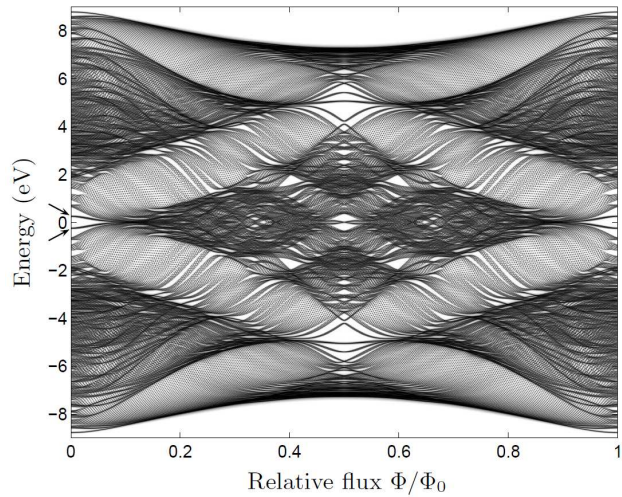}
\caption{
Hofstadter butterfly for the $\{12,8.6\}$ GAL, showing the density of states as function of energy and the relative flux through a graphene unit cell. To ease visibility, the logarithm of the density of states is plotted. The arrows indicate the states emerging due to regions of zigzag edges at the edge of the antidot.
}
\label{fig:L12R8p6}
\end{center}
\end{figure}

The Hofstadter butterflies for two examples of GALs, $\{4,2\}$ and $\{12, 8.6\}$, are illustrated in Figs.~\ref{fig:42GAL} and \ref{fig:L12R8p6}. The geometries of these GALs are illustrated in Fig.~\ref{fig:supercell}b.
Note that the radius is chosen such that we ensure that no dangling bonds are created. While these structures, having feature sizes below one nanometer, are smaller than what can be achieved experimentally, we will demonstrate below that certain scaling laws can be used to extrapolate results to GALs with more realistically sized features. Focusing first on the $\{4,2\}$ GAL we note that while the structure of the spectrum is significantly richer than that of gapped graphene in Fig.~\ref{fig:gapped}, many features of the gapped graphene spectrum are preserved. In general, the large regions devoid of eigenstates for gapped graphene tend to be connected by additional bands in the GAL case. This structure becomes increasingly rich as the size of the GAL unit cell is increased, as is evident in Fig.~\ref{fig:L12R8p6}. Comparing the two GALs, one significant difference is the additional states at zero magnetic field for the $\{12, 8.6\}$ GAL, marked by arrows in Fig.~\ref{fig:L12R8p6}. These emerge due to local regions of zigzag geometry at the edge of the hole, which tend to induce localized states.\cite{Gunst2011} The localized nature of the states residing in this band are reflected in its behaviour as the magnetic field is increased, where compared to other energy bands it remains very narrow with little splitting of the energy levels. 

\section{Magnetically induced band gap quenching}
\begin{figure}
\begin{center}
\includegraphics[width=.95\linewidth]{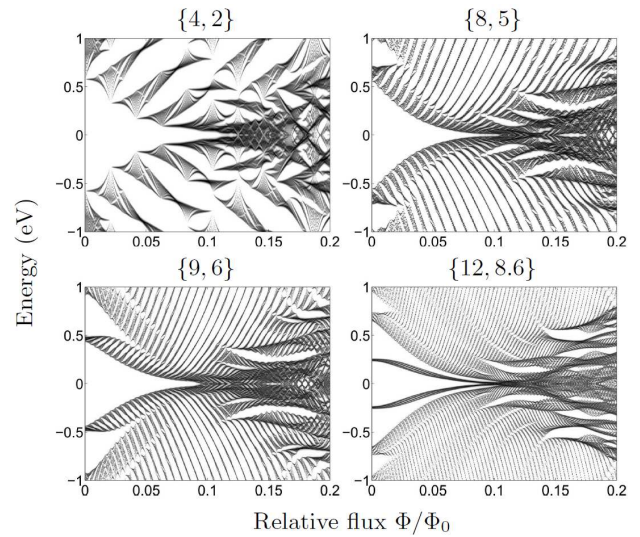}
\caption{
Closer view of the low-flux part of the Hofstadter butterflies for $\{4,2\}$, $\{8,5\}$, $\{9,6\}$ and $\{12,8.6\}$ GALs, showing the quenching of the gap by the magnetic field. Note that these GALs all have similar minimum neck widths.
}
\label{fig:42GALz}
\end{center}
\end{figure}
The most striking difference between the results for the GALs in Figs.~\ref{fig:42GAL} and \ref{fig:L12R8p6} and those of the simpler gapped graphene model shown in Fig.~\ref{fig:gapped} is the behaviour of the lowest Landau level as the magnetic field is increased. A Dirac treatment of gapped graphene predicts that the energy of this state should be $E_g/2$, independent of magnetic field, a behaviour that is confirmed in Fig.~\ref{fig:gapped} also for very large magnetic fields. However, the spectra for the GALs show an entirely different behaviour. Here, the band gap is quenched as the magnetic field strength is increased and eventually closes entirely. This is reminiscent of what is seen for armchair graphene nanoribbons\cite{Huang2007, Kumar2010} and graphene quantum dots.\cite{Zhang2008}
The crucial difference between GALs and a gapped graphene model is the additional characteristic lengths introduced by the antidots. In particular, we propose that the gap in a GAL will be quenched once the magnetic length $l_B$ is of the order of the minimum neck width of the GAL. In a simple picture, when the cyclotron radius becomes sufficiently small, the individual eigenstates do not sample the lattice sufficiently for the band gap to be fully resolved.
Because of the unique property of the lowest Landau level in bulk graphene, namely that it sits at the Dirac point energy regardless of magnetic field strength, this will result in a diminished band gap. We stress that we do not expect a similar effect to occur in 2DEGs based on semiconductor heterostructures, where the energy of the lowest Landau level is proportional to the magnetic field strength.

To illustrate the discussion above, we show in Fig.~\ref{fig:42GALz} zooms of the low-flux region of the spectra of different GALs, more clearly illustrating the magnetically induced band gap quenching. We show four examples of GALs, all of which have similar neck widths. Despite significant differences in the band gaps at zero magnetic field we find that the magnetic flux for which the gap is completely quenched is very similar for all four GALs. Note the appearance of additional very narrow bands at zero field in the case of the $\{9,6\}$ GALs (near $\pm 0.45$~eV) and $\{12,8.6\}$ GALs (near $\pm 0.25$~eV). As discussed above, these states appear due to quasi-localized states residing predominantly on regions of local zigzag geometry at the edge of the antidots. Ignoring these additional states for now, and taking the band gap to be defined as the gap between the wider energy bands, we see that the dependence of the band gap on magnetic field is very similar in all four cases. 

\begin{figure}
\begin{center}
\includegraphics[width=.9\linewidth]{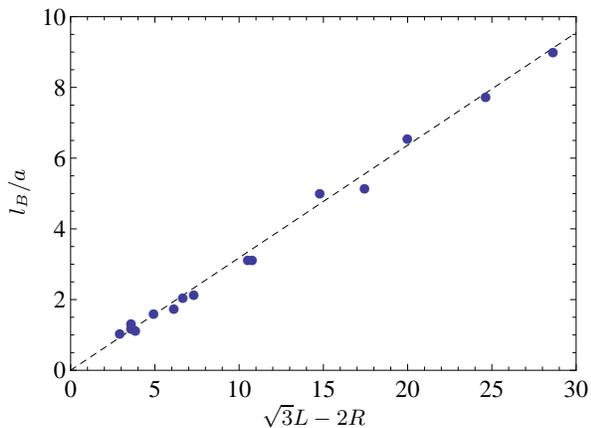}
\caption{(Color online)  The magnetic length $l_B$ corresponding to the critical flux for which the band gap is closed as a function of the neck width $w/a\simeq \sqrt{3}L-2R$ of the GAL. The dashed lines shows a best fit of the form $\alpha l_B=w$.
}
\label{fig:criticalFlux}
\end{center}
\end{figure}
In Fig.~\ref{fig:criticalFlux} we illustrate the geometry dependence of the magnetic length $l_B$ corresponding to the critical flux $\Phi_c$, for which the band gap is completely closed, on the minimum neck width. The neck width of a given GAL is approximately $w/a=\sqrt{3}L-2R$, an expression that is obviously more accurate for larger structures where the exact atomic details can be safely disregarded.
As expected from the physical picture described above, we find a clear linear dependence between the two quantities, such that the critical flux is defined via the simple relation $\alpha l_B=w$, with $\alpha\simeq 3.1$. 
Note that $l_B\simeq 25.7$~nm~T$^{1/2}/\sqrt{B}$, suggesting that the gap would be quenched at realistic magnetic field strengths for larger structures. Indeed, the scaling law provides us with a means of extrapolating our theoretical results to more realistically sized structures. As an example, we consider a GAL with a lattice constant of 60~nm and an antidot radius of 25~nm, which represent experimentally feasible feature sizes.\cite{Giebers2012} The neck width of this GAL is approximately $w\simeq 54$~nm, from which the scaling law predicts a critical flux of $\Phi_c\simeq 2.82\times 10^{-5}\Phi_0$. This translates to a magnetic field strength of $B\simeq 2.2$~T, well within reach of experiments.

\begin{figure}
\begin{center}
\includegraphics[width=\linewidth]{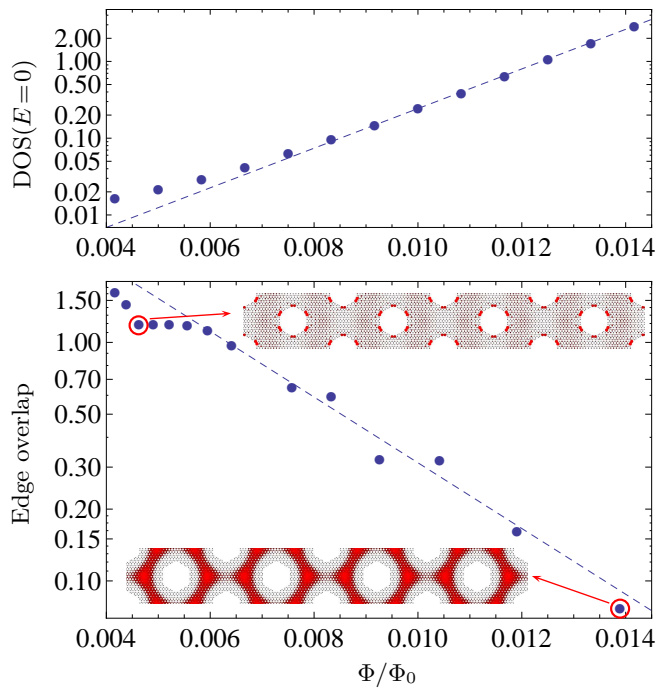}
\caption{(Color online) Magnetic flux dependence for a $\{12,5\}$ GAL of (upper panel) the density of states (in arbitrary units) at the Dirac point energy and (lower panel) the overlap with the antidot edge of the eigenstate corresponding to the lowest (positive) energy. The overlap is shown relative to the overlap assuming uniform distribution of the eigenstate. See the main text for details. In both panels, the dashed line shows best fits of exponential growth or decay, respectively. The insets in the lower panel show the probability density for the two eigenstates marked with circles.
}
\label{fig:OL}
\end{center}
\end{figure}
To further illustrate the interplay between the two length scales set by the magnetic field and the neck width of the GAL, we have studied the eigenstate nearest to the Dirac point energy. In particular, we consider the overlap $\int_\mathrm{edge}|\psi(x,y)|^2dxdy$ of the probability density with the edge of the antidot. In practice, we take this as being equal to sum over the absolute square $|c_n|^2$ of the expansion coefficients of the $\pi$-orbitals of carbon atoms with only two nearest neighbours. The lower panel of Fig.~\ref{fig:OL} shows the edge overlap as a function of the relative magnetic flux for a $\{12,5\}$ GAL. The edge overlap is shown relative to the overlap that would be expected if the probability density were evenly distributed across all carbon atoms, i.e. we plot $\sum_\mathrm{n\in\mathrm{edge}} |c_n|^2(N_\mathrm{tot}/N_\mathrm{edge})$, where $N_\mathrm{tot}$ is the total number of carbon atoms in the structure while $N_\mathrm{edge}$ is the number of carbon atoms with only two nearest neighbours. In the figure, we also include a best fit to the data, showing a clear exponential decay of the edge overlap with magnetic flux. For comparison, we show in the upper panel the density of states right at the Dirac point energy versus the magnetic flux. Here, we see an exponential increase of the DOS as the eigenstate is pushed away from the edge of the antidot. The critical flux where the band gap is closed corresponds quite accurately to the point where the exponential decay of the edge overlap starts to level off (for larger magnetic fields than included in the figure). In the insets of the lower panel of Fig.~\ref{fig:OL}, we show two charge densities corresponding to the magnetic flux and edge overlaps indicated with circles in the figure. Because of the high degeneracy of the eigenstates we show the probability densities summed over all states in the lowest band of energies. The black dots indicate carbon atoms while the size and shading of the filled, red circles illustrate the probability density. To ease visibility we show just the first four rectangular GAL unit cells, which correspond only to a small region of the full magnetic unit cell used in the calculations. The transition from a state with significant overlap with the antidot edge to a state localized predominantly between the antidots is quite evident.

These results fit very well with the physical picture, discussed above, of the band gap closing when the cyclotron radius becomes sufficiently small compared to the neck width of the GAL. It is also interesting to note the tendencies for the eigenstates to localize at the edge of the antidots for smaller magnetic fields, a result that fits well with previous studies of the magnetic properties of a single, isolated antidot in graphene.\cite{Isolated_antidot2012} 

\section{Summary and discussion}
Using a method based on the recursive Green's function formalism extended to deal with structures periodic in two dimensions, we have calculated Hofstadter butterflies of graphene antidot lattices (GALs). While the low-energy properties of GALs are usually well described in a simpler gapped graphene model, we find qualitative differences in the case of GALs in magnetic fields. In particular, the lowest Landau level is \emph{not} - as is the case in gapped graphene models - independent of magnetic field. Instead, we find that the GAL band gap can be effectively tuned by applying a magnetic field. In particular, the band gap is quenched entirely when the cyclotron radius becomes of the order of the neck width of the GAL. While for lower magnetic fields, the eigenstates nearest the Dirac point energy are localized predominantly at the antidot edges, we find that in the transition region, where the band gap is quenched, the states become increasingly localized between antidots. Using a simple scaling law we show that the predicted band gap quenching might be seen for reasonable magnetic field strengths in experimentally feasible structures.

We note that a similar effect of magnetically induced band gap quenching has been seen in graphene nanoribbons.\cite{Kumar2010} In the case of ribbons, however, the quenching occurs in a regime where the cyclotron radius is much larger than the nanoribbon width and, thus, before the formation of Landau levels. For GALs, we observe something quite different, namely a quenching of the gap that occurs for magnetic field strengths significantly larger than the onset of a Landau level structure. For ribbons, this effect has been used to predict an intrinsic magnetoresistive effect.\cite{Kumar2010} It is interesting to wonder whether very large negative magnetoresistance might be seen for GALs in magnetic fields due to the magnetically induced band gap quenching. Indeed, Giesbers \emph{et al.} have seen indications of negative magnetoresistance in measurements of GAL samples with features of the order 100~nm in magnetic fields in the range of a few Tesla.\cite{Giebers2012} Quite large negative magnetoresistance has also been seen at low temperature measurements of GALs with neck widths of roughly 50~nm in recently published results by Zhang \emph{et al.}\cite{Zhang2013} Our scaling result would suggest a quenching of the band gap at roughly $2.6$~T, which is of the same order of magnetic field that they see a significant increase in conductance. We note that we expect disorder to play a significant role in actual samples, the effect of which we have not considered in the present paper. However, an interesting point in relation to the results of Fig.~\ref{fig:OL} is that the tendency of the states to localize between the antidots might have the benefit of reducing any effects pertaining to the particular edge geometry of the hole. This would conceivably include any effects due to disorder on the edge of the antidot, which is to be expected in most realizations of GALs. We plan to pursuit further theoretical studies of these topics in future work.

\section{Acknowledgments}
The work by J.G.P. is financially supported by the Danish Council for Independent Research, FTP grant numbers
11-105204 and 11-120941. The Center for Nanostructured Graphene (CNG) is sponsored by the Danish National 
Research Foundation, Project DNRF58. We thank A.-P. Jauho for helpful comments during the development of the manuscript.

% ---------------------------------------------------------------------
% ---------------------------------------------------------------------
% APPENDIX
% ---------------------------------------------------------------------
% ---------------------------------------------------------------------

\appendix
\section{Recursive method}
Our numerical method is based on standard recursive Green's function techniques, wherein advantage is taken of the block-diagonal form of the Hamiltonian.
Here, we briefly outline the standard formalism and extend it to incorporate periodic boundary conditions in both directions.
\begin{figure}
\begin{center}
\includegraphics[width=\linewidth]{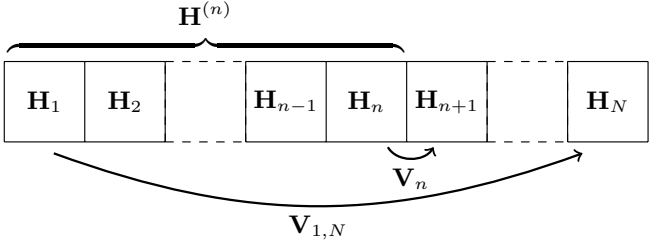}
\caption{The Hamiltonian $\mathbf{H}^{(N)}$ of the complete system can be assembled from Hamiltonians of isolated subsystems $\mathbf{H}_n$ and the matrices $\mathbf{V}_n$ describing the coupling between the subsystems. Note that due to periodicity, the first and final subsystems are coupled via the matrix $\mathbf{V}_{1,N}$.}
\label{fig:rgf}
\end{center}
\end{figure}
We consider the whole system as being assembled via $N$ subsystems, as illustrated in Fig.~\ref{fig:rgf}. Each isolated subsystem is characterized by a Hamiltonian
matrix $\mathbf{H}_n$ and a corresponding Green's function $\mathbf{G}_n = [(E+i\gamma)\mathbf{I} - \mathbf{H}_n]^{-1}$, with $\mathbf{I}$ an identity matrix, and where we have included the broadening term $\gamma$.
The coupling between the $n$'th and the $n+1$'th subsystem is described by the matrix $\mathbf{V}_{n}$. Further, we define $\mathbf{H}^{(n)}$ and $\mathbf{G}^{(n)}$ as the Hamiltonian and Green's function, respectively, of the combined system of the first $n$ subsystems, including the coupling between them. 
Taking advantage of the block-diagonal form of $\mathbf{H}^{(n)}$ for $n<N$, the standard recursive Green's function method relies on a recursive procedure
for $n+1<N$ reading
\begin{eqnarray}
\GGG{n+1}{n+1,n+1} &\!\!=\!\!& 
\left(\left(E+i\gamma\right)\mathbf{I}-\HHH_{n+1}-\boldsymbol{\Sigma}_n \right)^{-1} \nonumber \\
\GGG{n+1}{i,n+1} &\!\!=\!\!& 
\GGG{n}{i,n}\VVV_{n}\GGG{n+1}{n+1,n+1} \;\;\textrm{, } i\leq n, \nonumber \\
\GGG{n+1}{n+1,j} &\!\!=\!\!& 
\GGG{n+1}{n+1,n+1}\VVV_{n}^\dagger\GGG{n}{n,j} \;\;\textrm{, } j\leq n, \nonumber \\
\GGG{n+1}{i,j} &\!\!=\!\!& 
\GGG{n}{i,j}+
\GGG{n}{i,n}\VVV_{n}\GGG{n+1}{n+1,n+1}\VVV_{n}^\dagger\GGG{n}{n,j}\;\;\textrm{, } i,j \leq n,\nonumber \\
\label{eq:Grec}
\end{eqnarray}
where we have introduced the self-energy
$\boldsymbol{\Sigma}_n = \VVV_n^\dagger\GGG{n}{n,n}\VVV_n$, which takes care of the coupling between the subsystems.
Note that the subscripts of the Green's functions refer not to individual elements but to submatrices defined by the sizes of the
subsystems.
For the last step of the recursive procedure, $n+1=N$, we do take into account the coupling to the first subsystem due to
periodicity, resulting in
\begin{eqnarray}
\GGG{N}{N,N} &\!\!=\!\!& 
\left((E+i\gamma)\mathbf{I}-\HHH_{N}-\boldsymbol{\Sigma}_{N\!-\!1} \right)^{-1} \nonumber \\
\GGG{N}{i,N} &\!\!=\!\!& 
\left(\GGG{N\!-\!1}{i,1}\VVV_{1,N}+\GGG{N\!-\!1}{i,N\!-\!1}\VVV_{N\!-\!1}\right)\GGG{N}{N,N} \;\;\textrm{, } i<N, \nonumber \\
\GGG{N}{N,j} &\!\!=\!\!& 
\GGG{N}{N,N}\left(\VVV_{1,N}^\dagger\GGG{N\!-\!1}{1,j}+\VVV_{N\!-\!1}^\dagger\GGG{N\!-\!1}{N\!-\!1,j}\right) \;\;\textrm{, } j<N, \nonumber \\
\GGG{N}{i,j} &\!\!=\!\!& 
\GGG{N\!-\!1}{i,j}+
\left(\GGG{N\!-\!1}{i,1}\VVV_{1,N}+\GGG{N\!-\!1}{i,N\!-\!1}\VVV_{N\!-\!1}\right)
\GGG{N}{N,N}
\nonumber \\
&&\times
\left(\VVV_{1,N}^\dagger\GGG{N\!-\!1}{1,j}+\VVV_{N\!-\!1}^\dagger\GGG{N\!-\!1}{N\!-\!1,j}\right)
\;\;\textrm{, } i,j < N,\nonumber \\
\label{eq:GrecN}
\end{eqnarray}
with the self-energy
\begin{eqnarray}
\boldsymbol{\Sigma}_{N-1} &=& 
\left(\VVV_{1,N}^\dagger\GGG{N\!-\!1}{1,N\!-\!1}+\VVV_{N\!-\!1}^\dagger\GGG{N\!-\!1}{N\!-\!1,N\!-\!1}\right)\VVV_{N\!-\!1}
\nonumber \\
&&+ 
\left(\VVV_{1,N}^\dagger\GGG{N\!-\!1}{1,1}+\VVV_{N\!-\!1}^\dagger\GGG{N\!-\!1}{N\!-\!1,1}\right)\VVV_{1,N}.
\end{eqnarray}

While the method above can be used directly to obtain the full Green's function $\mathbf{G}^{(N)}$ of the whole system, when dealing with very large systems this method is both too slow and too memory intensive. Instead, we derive a recursive algorithm specifically for calculating the density of states. We calculate the density of states via the relation
$\rho(E) = -\pi^{-1}\mathrm{Im}\{T_G^{(N)}\}$, where we have defined $T_G^{(n)} = \mathrm{Tr}[\mathbf{G}^{(n)}]$.
We first treat the case where $n+1<N$ or periodicity is ignored.
Using the relations for the elements of the Green's function given in Eqs.~(\ref{eq:Grec}), and doing a bit of algebra, we arrive at
\begin{equation}
T_G^{(n+1)} = T_G^{(n)}+\mathrm{Tr}\left[\GnP\left(\VVV_n^\dagger\mathbf{E}_\mathrm{nn}^{(n)}\VVV_n+\mathbf{I}\right)\right],
\end{equation}
where to ease notation we have defined the shorthand $\GnP = \GGG{n+1}{n+1,n+1}$. Furthermore, we have introduced the ancillary matrix
\begin{equation}
\mathbf{E}_\mathrm{nn}^{(n)} = \sum_i^n \GGG{n}{n,i}\GGG{n}{i,n},
\end{equation}
with the recursive relation
\begin{equation}
\mathbf{E}_\mathrm{nn}^{(n+1)} = \GnP\left(\VVV_n^\dagger \mathbf{E}_\mathrm{nn}^{(n)}\VVV_n+\mathbf{I}\right)\GnP.
\end{equation}
While our notation is slightly different, we note that this result has previously been derived by MacKinnon.\cite{MacKinnon1980, MacKinnon1985} We include the results here for completeness.
For the final step of the recursive algorithm we take into account periodicity of the structure via Eqs.~(\ref{eq:GrecN}) which leads to
%\begin{widetext}
\begin{eqnarray}
T_G^{(N)} &\!=\!& T_G^{(N\!-\!1)} \!+\! \mathrm{Tr}\left[
\mathbf{G}_N\left(
\VVV_{1,N}^\dagger\Enm{N\!-\!1}{11}\VVV_{1,N}
\right.\right.\nonumber \\
&& \left.\left.
+
\VVV_{1,N}^\dagger\Enm{N\!-\!1}{1n}\VVV_{N\!-\!1}
+
\VVV_{N\!-\!1}^\dagger\Enm{N\!-\!1}{n1}\VVV_{1,N}
\right.\right.\nonumber \\
&& \left.\left.
+
\VVV_{N\!-\!1}^\dagger\Enm{N\!-\!1}{nn}\VVV_{N\!-\!1}+
\mathbf{I}
\right)\!
\right],
\end{eqnarray}
%\end{widetext}
where we have introduced additional ancillary matrices
\begin{eqnarray}
\Enm{n}{n1} &=& \sum_i^{n}\GGG{n}{n,i}\GGG{n}{i,1},\\
\Enm{n}{1n} &=& \sum_i^{n}\GGG{n}{1,i}\GGG{n}{i,n},\\
\Enm{n}{11} &=& \sum_i^{n}\GGG{n}{1,i}\GGG{n}{i,1},
\end{eqnarray}
with recursive relations
%\begin{widetext}
\begin{eqnarray}
\!\!\!\!\!\!\!\!\!\!\!\Enm{n+1}{n1}
&\!\!\!\!=\!\!\!&
\GnP\!\!\left[
\VVV_n^\dagger\Enm{n}{n1} \!\!+\!\! \left(\VVV_n^\dagger\Enm{n}{nn}\VVV_n\!\!+\!\!\mathbf{I}\right)\GGG{n+1}{n+1,1}
\right],
\end{eqnarray}
and similarly
\begin{eqnarray}
\!\!\!\!\!\!\!\!\!\!\!\Enm{n+1}{1n}
&\!\!\!\!=\!\!\!&
\left[
\Enm{n}{1n}\VVV_n \!\!+\!\! \GGG{n+1}{1,n+1}\left(\VVV_n^\dagger\Enm{n}{nn}\VVV_n\!\!+\!\!\mathbf{I}\right)
\right]\!\!
\GnP.
\end{eqnarray}
Finally, we have
\begin{eqnarray}
\Enm{n+1}{11}
&=& \Enm{n}{11} 
+ \GGG{n+1}{1,n+1}\VVV_n^\dagger\Enm{n}{n1}
+ \Enm{n}{1n}\VVV_n\GGG{n+1}{n+1,1}
\nonumber \\
&&
+ \GGG{n+1}{1,n+1}\left(\VVV_n^\dagger\Enm{n}{nn}\VVV_n+\mathbf{I}\right)\GGG{n+1}{n+1,1},
\end{eqnarray}
%\end{widetext}
while the recursive procedures for $\GGG{n}{1,n}$ and $\GGG{n}{n,1}$ are of course given by Eqs.~(\ref{eq:Grec}).


\begin{thebibliography}{42}
\expandafter\ifx\csname natexlab\endcsname\relax\def\natexlab#1{#1}\fi
\expandafter\ifx\csname bibnamefont\endcsname\relax
  \def\bibnamefont#1{#1}\fi
\expandafter\ifx\csname bibfnamefont\endcsname\relax
  \def\bibfnamefont#1{#1}\fi
\expandafter\ifx\csname citenamefont\endcsname\relax
  \def\citenamefont#1{#1}\fi
\expandafter\ifx\csname url\endcsname\relax
  \def\url#1{\texttt{#1}}\fi
\expandafter\ifx\csname urlprefix\endcsname\relax\def\urlprefix{URL }\fi
\providecommand{\bibinfo}[2]{#2}
\providecommand{\eprint}[2][]{\url{#2}}

\bibitem[{\citenamefont{Weiss et~al.}(1993)\citenamefont{Weiss, Richter,
  Menschig, Bergmann, Schweizer, von Klitzing, and
  Weimann}}]{PhysRevLett.70.4118}
\bibinfo{author}{\bibfnamefont{D.}~\bibnamefont{Weiss}},
  \bibinfo{author}{\bibfnamefont{K.}~\bibnamefont{Richter}},
  \bibinfo{author}{\bibfnamefont{A.}~\bibnamefont{Menschig}},
  \bibinfo{author}{\bibfnamefont{R.}~\bibnamefont{Bergmann}},
  \bibinfo{author}{\bibfnamefont{H.}~\bibnamefont{Schweizer}},
  \bibinfo{author}{\bibfnamefont{K.}~\bibnamefont{von Klitzing}},
  \bibnamefont{and} \bibinfo{author}{\bibfnamefont{G.}~\bibnamefont{Weimann}},
  \bibinfo{journal}{Phys. Rev. Lett.} \textbf{\bibinfo{volume}{70}},
  \bibinfo{pages}{4118} (\bibinfo{year}{1993}).

\bibitem[{\citenamefont{Ensslin and Petroff}(1990)}]{PhysRevB.41.12307}
\bibinfo{author}{\bibfnamefont{K.}~\bibnamefont{Ensslin}} \bibnamefont{and}
  \bibinfo{author}{\bibfnamefont{P.~M.} \bibnamefont{Petroff}},
  \bibinfo{journal}{Phys. Rev. B} \textbf{\bibinfo{volume}{41}},
  \bibinfo{pages}{12307} (\bibinfo{year}{1990}).

\bibitem[{\citenamefont{Pedersen
  et~al.}(2008{\natexlab{a}})\citenamefont{Pedersen, Flindt, Pedersen,
  Mortensen, Jauho, and Pedersen}}]{ar_tgp1}
\bibinfo{author}{\bibfnamefont{T.~G.} \bibnamefont{Pedersen}},
  \bibinfo{author}{\bibfnamefont{C.}~\bibnamefont{Flindt}},
  \bibinfo{author}{\bibfnamefont{J.}~\bibnamefont{Pedersen}},
  \bibinfo{author}{\bibfnamefont{N.~A.} \bibnamefont{Mortensen}},
  \bibinfo{author}{\bibfnamefont{A.-P.} \bibnamefont{Jauho}}, \bibnamefont{and}
  \bibinfo{author}{\bibfnamefont{K.}~\bibnamefont{Pedersen}},
  \bibinfo{journal}{Phys. Rev. Lett.} \textbf{\bibinfo{volume}{100}},
  \bibinfo{pages}{136804} (\bibinfo{year}{2008}{\natexlab{a}}).

\bibitem[{\citenamefont{Pedersen
  et~al.}(2008{\natexlab{b}})\citenamefont{Pedersen, Flindt, Pedersen, Jauho,
  Mortensen, and Pedersen}}]{ar_tgp2}
\bibinfo{author}{\bibfnamefont{T.~G.} \bibnamefont{Pedersen}},
  \bibinfo{author}{\bibfnamefont{C.}~\bibnamefont{Flindt}},
  \bibinfo{author}{\bibfnamefont{J.}~\bibnamefont{Pedersen}},
  \bibinfo{author}{\bibfnamefont{A.~P.} \bibnamefont{Jauho}},
  \bibinfo{author}{\bibfnamefont{N.~A.} \bibnamefont{Mortensen}},
  \bibnamefont{and} \bibinfo{author}{\bibfnamefont{K.}~\bibnamefont{Pedersen}},
  \bibinfo{journal}{Phys. Rev. B} \textbf{\bibinfo{volume}{77}},
  \bibinfo{pages}{245431} (\bibinfo{year}{2008}{\natexlab{b}}).

\bibitem[{\citenamefont{Shen et~al.}(2008)\citenamefont{Shen, Wu, Capano,
  Rokhinson, Engel, and Ye}}]{shen2008}
\bibinfo{author}{\bibfnamefont{T.}~\bibnamefont{Shen}},
  \bibinfo{author}{\bibfnamefont{Y.~Q.} \bibnamefont{Wu}},
  \bibinfo{author}{\bibfnamefont{M.~A.} \bibnamefont{Capano}},
  \bibinfo{author}{\bibfnamefont{L.~P.} \bibnamefont{Rokhinson}},
  \bibinfo{author}{\bibfnamefont{L.~W.} \bibnamefont{Engel}}, \bibnamefont{and}
  \bibinfo{author}{\bibfnamefont{P.~D.} \bibnamefont{Ye}},
  \bibinfo{journal}{Appl. Phys. Lett.} \textbf{\bibinfo{volume}{93}},
  \bibinfo{eid}{122102} (\bibinfo{year}{2008}).

\bibitem[{\citenamefont{Eroms and Weiss}(2009)}]{Eroms2009}
\bibinfo{author}{\bibfnamefont{J.}~\bibnamefont{Eroms}} \bibnamefont{and}
  \bibinfo{author}{\bibfnamefont{D.}~\bibnamefont{Weiss}},
  \bibinfo{journal}{New J. Phys.} \textbf{\bibinfo{volume}{11}},
  \bibinfo{pages}{095021} (\bibinfo{year}{2009}).

\bibitem[{\citenamefont{Shimizu et~al.}(2012)\citenamefont{Shimizu, Nakamura,
  Tada, Yagi, and Haruyama}}]{Shimizu2012}
\bibinfo{author}{\bibfnamefont{T.}~\bibnamefont{Shimizu}},
  \bibinfo{author}{\bibfnamefont{J.}~\bibnamefont{Nakamura}},
  \bibinfo{author}{\bibfnamefont{K.}~\bibnamefont{Tada}},
  \bibinfo{author}{\bibfnamefont{Y.}~\bibnamefont{Yagi}}, \bibnamefont{and}
  \bibinfo{author}{\bibfnamefont{J.}~\bibnamefont{Haruyama}},
  \bibinfo{journal}{Appl. Phys. Lett.} \textbf{\bibinfo{volume}{100}},
  \bibinfo{eid}{023104} (\bibinfo{year}{2012}).

\bibitem[{\citenamefont{Giesbers et~al.}(2012)\citenamefont{Giesbers, Peters,
  Burghard, and Kern}}]{Giebers2012}
\bibinfo{author}{\bibfnamefont{A.~J.~M.} \bibnamefont{Giesbers}},
  \bibinfo{author}{\bibfnamefont{E.~C.} \bibnamefont{Peters}},
  \bibinfo{author}{\bibfnamefont{M.}~\bibnamefont{Burghard}}, \bibnamefont{and}
  \bibinfo{author}{\bibfnamefont{K.}~\bibnamefont{Kern}},
  \bibinfo{journal}{Phys. Rev. B} \textbf{\bibinfo{volume}{86}},
  \bibinfo{pages}{045445} (\bibinfo{year}{2012}).

\bibitem[{\citenamefont{Kim et~al.}(2010)\citenamefont{Kim, Safron, Han,
  Arnold, and Gopalan}}]{Kim2010}
\bibinfo{author}{\bibfnamefont{M.}~\bibnamefont{Kim}},
  \bibinfo{author}{\bibfnamefont{N.~S.} \bibnamefont{Safron}},
  \bibinfo{author}{\bibfnamefont{E.}~\bibnamefont{Han}},
  \bibinfo{author}{\bibfnamefont{M.~S.} \bibnamefont{Arnold}},
  \bibnamefont{and} \bibinfo{author}{\bibfnamefont{P.}~\bibnamefont{Gopalan}},
  \bibinfo{journal}{Nano Lett.} \textbf{\bibinfo{volume}{10}},
  \bibinfo{pages}{1125} (\bibinfo{year}{2010}).

\bibitem[{\citenamefont{Bai et~al.}(2010)\citenamefont{Bai, Zhong, Jiang,
  Huang, and Duan}}]{Bai2010}
\bibinfo{author}{\bibfnamefont{J.}~\bibnamefont{Bai}},
  \bibinfo{author}{\bibfnamefont{X.}~\bibnamefont{Zhong}},
  \bibinfo{author}{\bibfnamefont{S.}~\bibnamefont{Jiang}},
  \bibinfo{author}{\bibfnamefont{Y.}~\bibnamefont{Huang}}, \bibnamefont{and}
  \bibinfo{author}{\bibfnamefont{X.}~\bibnamefont{Duan}},
  \bibinfo{journal}{Nat. Nanotechn.} \textbf{\bibinfo{volume}{5}},
  \bibinfo{pages}{190} (\bibinfo{year}{2010}).

\bibitem[{\citenamefont{Balog et~al.}(2010)\citenamefont{Balog, J{\o}rgensen,
  Nilsson, Andersen, Rienks, Bianchi, Fanetti, L{\ae}gsgaard, Baraldi, Lizzit
  et~al.}}]{ar_Balog}
\bibinfo{author}{\bibfnamefont{R.}~\bibnamefont{Balog}},
  \bibinfo{author}{\bibfnamefont{B.}~\bibnamefont{J{\o}rgensen}},
  \bibinfo{author}{\bibfnamefont{L.}~\bibnamefont{Nilsson}},
  \bibinfo{author}{\bibfnamefont{M.}~\bibnamefont{Andersen}},
  \bibinfo{author}{\bibfnamefont{E.}~\bibnamefont{Rienks}},
  \bibinfo{author}{\bibfnamefont{M.}~\bibnamefont{Bianchi}},
  \bibinfo{author}{\bibfnamefont{M.}~\bibnamefont{Fanetti}},
  \bibinfo{author}{\bibfnamefont{E.}~\bibnamefont{L{\ae}gsgaard}},
  \bibinfo{author}{\bibfnamefont{A.}~\bibnamefont{Baraldi}},
  \bibinfo{author}{\bibfnamefont{S.}~\bibnamefont{Lizzit}},
  \bibnamefont{et~al.}, \bibinfo{journal}{Nat. Mater.}
  \textbf{\bibinfo{volume}{9}}, \bibinfo{pages}{315} (\bibinfo{year}{2010}).

\bibitem[{\citenamefont{Xu et~al.}(2013)\citenamefont{Xu, Wu, Schneider,
  Houben, Malladi, Dekker, Yucelen, Dunin-Borkowski, and Zandbergen}}]{Xu2013}
\bibinfo{author}{\bibfnamefont{Q.}~\bibnamefont{Xu}},
  \bibinfo{author}{\bibfnamefont{M.-Y.} \bibnamefont{Wu}},
  \bibinfo{author}{\bibfnamefont{G.~F.} \bibnamefont{Schneider}},
  \bibinfo{author}{\bibfnamefont{L.}~\bibnamefont{Houben}},
  \bibinfo{author}{\bibfnamefont{S.~K.} \bibnamefont{Malladi}},
  \bibinfo{author}{\bibfnamefont{C.}~\bibnamefont{Dekker}},
  \bibinfo{author}{\bibfnamefont{E.}~\bibnamefont{Yucelen}},
  \bibinfo{author}{\bibfnamefont{R.~E.} \bibnamefont{Dunin-Borkowski}},
  \bibnamefont{and} \bibinfo{author}{\bibfnamefont{H.~W.}
  \bibnamefont{Zandbergen}}, \bibinfo{journal}{ACS Nano}
  (\bibinfo{year}{2013}).

\bibitem[{\citenamefont{Park et~al.}(2010)\citenamefont{Park, Kim, and
  Yang}}]{Park2010}
\bibinfo{author}{\bibfnamefont{P.~S.} \bibnamefont{Park}},
  \bibinfo{author}{\bibfnamefont{S.~C.} \bibnamefont{Kim}}, \bibnamefont{and}
  \bibinfo{author}{\bibfnamefont{S.-R.~E.} \bibnamefont{Yang}},
  \bibinfo{journal}{J. Phys.: Condens. Mat.} \textbf{\bibinfo{volume}{22}},
  \bibinfo{pages}{375302} (\bibinfo{year}{2010}).

\bibitem[{\citenamefont{Kim and Yang}(2012)}]{Kim2012}
\bibinfo{author}{\bibfnamefont{S.~C.} \bibnamefont{Kim}} \bibnamefont{and}
  \bibinfo{author}{\bibfnamefont{S.-R.~E.} \bibnamefont{Yang}},
  \bibinfo{journal}{J. Phys.: Condens. Mat.} \textbf{\bibinfo{volume}{24}},
  \bibinfo{pages}{195301} (\bibinfo{year}{2012}).

\bibitem[{\citenamefont{Pedersen and
  Pedersen}(2012{\natexlab{a}})}]{Isolated_antidot2012}
\bibinfo{author}{\bibfnamefont{J.~G.} \bibnamefont{Pedersen}} \bibnamefont{and}
  \bibinfo{author}{\bibfnamefont{T.~G.} \bibnamefont{Pedersen}},
  \bibinfo{journal}{Phys. Rev. B} \textbf{\bibinfo{volume}{85}},
  \bibinfo{pages}{035413} (\bibinfo{year}{2012}{\natexlab{a}}).

\bibitem[{\citenamefont{Hofstadter}(1976)}]{Hofstadter1976}
\bibinfo{author}{\bibfnamefont{D.~R.} \bibnamefont{Hofstadter}},
  \bibinfo{journal}{Phys. Rev. B} \textbf{\bibinfo{volume}{14}},
  \bibinfo{pages}{2239} (\bibinfo{year}{1976}).

\bibitem[{\citenamefont{Pedersen}(2003)}]{Faraday_graphite}
\bibinfo{author}{\bibfnamefont{T.~G.} \bibnamefont{Pedersen}},
  \bibinfo{journal}{Phys. Rev. B} \textbf{\bibinfo{volume}{68}},
  \bibinfo{pages}{245104} (\bibinfo{year}{2003}).

\bibitem[{\citenamefont{Hasegawa and Kohmoto}(2006)}]{Hasegawa2006}
\bibinfo{author}{\bibfnamefont{Y.}~\bibnamefont{Hasegawa}} \bibnamefont{and}
  \bibinfo{author}{\bibfnamefont{M.}~\bibnamefont{Kohmoto}},
  \bibinfo{journal}{Phys. Rev. B} \textbf{\bibinfo{volume}{74}},
  \bibinfo{pages}{155415} (\bibinfo{year}{2006}).

\bibitem[{\citenamefont{Nemec and Cuniberti}(2007)}]{Nemec2007}
\bibinfo{author}{\bibfnamefont{N.}~\bibnamefont{Nemec}} \bibnamefont{and}
  \bibinfo{author}{\bibfnamefont{G.}~\bibnamefont{Cuniberti}},
  \bibinfo{journal}{Phys. Rev. B} \textbf{\bibinfo{volume}{75}},
  \bibinfo{pages}{201404} (\bibinfo{year}{2007}).

\bibitem[{\citenamefont{Wang et~al.}(2012)\citenamefont{Wang, Liu, and
  Chou}}]{Wang2012}
\bibinfo{author}{\bibfnamefont{Z.~F.} \bibnamefont{Wang}},
  \bibinfo{author}{\bibfnamefont{F.}~\bibnamefont{Liu}}, \bibnamefont{and}
  \bibinfo{author}{\bibfnamefont{M.~Y.} \bibnamefont{Chou}},
  \bibinfo{journal}{Nano Lett.} \textbf{\bibinfo{volume}{12}},
  \bibinfo{pages}{3833} (\bibinfo{year}{2012}).

\bibitem[{\citenamefont{Islamoglu, Oktel, and
  G\"ulseren}}]{Islamoglu2012}
\bibinfo{author}{\bibfnamefont{S.}~\bibnamefont{Islamoglu}},
  \bibinfo{author}{\bibfnamefont{M.~O.} \bibnamefont{Oktel}}, \bibnamefont{and}
  \bibinfo{author}{\bibfnamefont{O.~} \bibnamefont{G\"ulseren}},
  \bibinfo{journal}{Phys. Rev. B} \textbf{\bibinfo{volume}{85}},
  \bibinfo{pages}{235414} (\bibinfo{year}{2012}).

\bibitem[{\citenamefont{Zhang et~al.}(2008)\citenamefont{Zhang, Chang, and
  Peeters}}]{Zhang2008}
\bibinfo{author}{\bibfnamefont{Z.~Z.} \bibnamefont{Zhang}},
  \bibinfo{author}{\bibfnamefont{K.}~\bibnamefont{Chang}}, \bibnamefont{and}
  \bibinfo{author}{\bibfnamefont{F.~M.} \bibnamefont{Peeters}},
  \bibinfo{journal}{Phys. Rev. B} \textbf{\bibinfo{volume}{77}},
  \bibinfo{pages}{235411} (\bibinfo{year}{2008}).

\bibitem[{\citenamefont{Dean et~al.}(2012)\citenamefont{Dean, Wang, Maher,
  Forsythe, Ghahari, Gao, Katoch, Ishigami, Moon, Koshino et~al.}}]{Dean2012}
\bibinfo{author}{\bibfnamefont{C.~R.} \bibnamefont{Dean}},
  \bibinfo{author}{\bibfnamefont{L.}~\bibnamefont{Wang}},
  \bibinfo{author}{\bibfnamefont{P.}~\bibnamefont{Maher}},
  \bibinfo{author}{\bibfnamefont{C.}~\bibnamefont{Forsythe}},
  \bibinfo{author}{\bibfnamefont{F.}~\bibnamefont{Ghahari}},
  \bibinfo{author}{\bibfnamefont{Y.}~\bibnamefont{Gao}},
  \bibinfo{author}{\bibfnamefont{J.}~\bibnamefont{Katoch}},
  \bibinfo{author}{\bibfnamefont{M.}~\bibnamefont{Ishigami}},
  \bibinfo{author}{\bibfnamefont{P.}~\bibnamefont{Moon}},
  \bibinfo{author}{\bibfnamefont{M.}~\bibnamefont{Koshino}},
  \bibnamefont{et~al.}, \bibinfo{journal}{arXiv:1212.4783}
  (\bibinfo{year}{2012}).

\bibitem[{\citenamefont{Ponomarenko et~al.}(2012)\citenamefont{Ponomarenko,
  Gorbachev, Elias, Yu, Mayorov, Wallbank, Mucha-Kruczynski, Patel, Piot,
  Potemski et~al.}}]{Ponomarenko2012}
\bibinfo{author}{\bibfnamefont{L.~A.} \bibnamefont{Ponomarenko}},
  \bibinfo{author}{\bibfnamefont{R.~V.} \bibnamefont{Gorbachev}},
  \bibinfo{author}{\bibfnamefont{D.~C.} \bibnamefont{Elias}},
  \bibinfo{author}{\bibfnamefont{G.~L.} \bibnamefont{Yu}},
  \bibinfo{author}{\bibfnamefont{A.~S.} \bibnamefont{Mayorov}},
  \bibinfo{author}{\bibfnamefont{J.}~\bibnamefont{Wallbank}},
  \bibinfo{author}{\bibfnamefont{M.}~\bibnamefont{Mucha-Kruczynski}},
  \bibinfo{author}{\bibfnamefont{A.}~\bibnamefont{Patel}},
  \bibinfo{author}{\bibfnamefont{B.~A.} \bibnamefont{Piot}},
  \bibinfo{author}{\bibfnamefont{M.}~\bibnamefont{Potemski}},
  \bibnamefont{et~al.}, \bibinfo{journal}{arXiv:1212.5012}
  (\bibinfo{year}{2012}).

\bibitem[{\citenamefont{Castro et~al.}(2007)\citenamefont{Castro, Novoselov,
  Morozov, Peres, dos Santos, Nilsson, Guinea, Geim, and Neto}}]{Castro2007}
\bibinfo{author}{\bibfnamefont{E.~V.} \bibnamefont{Castro}},
  \bibinfo{author}{\bibfnamefont{K.~S.} \bibnamefont{Novoselov}},
  \bibinfo{author}{\bibfnamefont{S.~V.} \bibnamefont{Morozov}},
  \bibinfo{author}{\bibfnamefont{N.~M.~R.} \bibnamefont{Peres}},
  \bibinfo{author}{\bibfnamefont{J.~M.~B.} \bibnamefont{Lopes dos Santos}},
  \bibinfo{author}{\bibfnamefont{J.}~\bibnamefont{Nilsson}},
  \bibinfo{author}{\bibfnamefont{F.}~\bibnamefont{Guinea}},
  \bibinfo{author}{\bibfnamefont{A.~K.} \bibnamefont{Geim}}, \bibnamefont{and}
  \bibinfo{author}{\bibfnamefont{A.~H.} \bibnamefont{Castro~Neto}},
  \bibinfo{journal}{Phys. Rev. Lett.} \textbf{\bibinfo{volume}{99}},
  \bibinfo{pages}{216802} (\bibinfo{year}{2007}).

\bibitem[{\citenamefont{Zhang et~al.}(2010)\citenamefont{Zhang, Dai, Shi, Feng,
  and Zhang}}]{Zhang2010}
\bibinfo{author}{\bibfnamefont{A.}~\bibnamefont{Zhang}},
  \bibinfo{author}{\bibfnamefont{Z.}~\bibnamefont{Dai}},
  \bibinfo{author}{\bibfnamefont{L.}~\bibnamefont{Shi}},
  \bibinfo{author}{\bibfnamefont{Y.~P.} \bibnamefont{Feng}}, \bibnamefont{and}
  \bibinfo{author}{\bibfnamefont{C.}~\bibnamefont{Zhang}}, \bibinfo{journal}{J.
  Chem. Phys.} \textbf{\bibinfo{volume}{133}}, \bibinfo{pages}{224705}
  (\bibinfo{year}{2010}).

\bibitem[{\citenamefont{Pedersen and
  Pedersen}(2012{\natexlab{b}})}]{Pedersen2012}
\bibinfo{author}{\bibfnamefont{J.~G.} \bibnamefont{Pedersen}} \bibnamefont{and}
  \bibinfo{author}{\bibfnamefont{T.~G.} \bibnamefont{Pedersen}},
  \bibinfo{journal}{Phys. Rev. B} \textbf{\bibinfo{volume}{85}},
  \bibinfo{pages}{235432} (\bibinfo{year}{2012}{\natexlab{b}}).

\bibitem[{\citenamefont{Pedersen et~al.}(2009)\citenamefont{Pedersen, Jauho,
  and Pedersen}}]{Gapped2009}
\bibinfo{author}{\bibfnamefont{T.~G.} \bibnamefont{Pedersen}},
  \bibinfo{author}{\bibfnamefont{A.-P.} \bibnamefont{Jauho}}, \bibnamefont{and}
  \bibinfo{author}{\bibfnamefont{K.}~\bibnamefont{Pedersen}},
  \bibinfo{journal}{Phys. Rev. B} \textbf{\bibinfo{volume}{79}},
  \bibinfo{pages}{113406} (\bibinfo{year}{2009}).

\bibitem[{\citenamefont{Gusynin et~al.}(2007)\citenamefont{Gusynin, Sharapov,
  and Carbotte}}]{Gusynin2007}
\bibinfo{author}{\bibfnamefont{V.~P.} \bibnamefont{Gusynin}},
  \bibinfo{author}{\bibfnamefont{S.~G.} \bibnamefont{Sharapov}},
  \bibnamefont{and} \bibinfo{author}{\bibfnamefont{J.~P.}
  \bibnamefont{Carbotte}}, \bibinfo{journal}{J. Phys.: Condens. Mat.}
  \textbf{\bibinfo{volume}{19}}, \bibinfo{pages}{026222}
  (\bibinfo{year}{2007}).

\bibitem[{\citenamefont{Pedersen and Pedersen}(2011)}]{Magneto_gapped2011}
\bibinfo{author}{\bibfnamefont{J.~G.} \bibnamefont{Pedersen}} \bibnamefont{and}
  \bibinfo{author}{\bibfnamefont{T.~G.} \bibnamefont{Pedersen}},
  \bibinfo{journal}{Phys. Rev. B} \textbf{\bibinfo{volume}{84}},
  \bibinfo{pages}{115424} (\bibinfo{year}{2011}).

\bibitem[{\citenamefont{Huang et~al.}(2007)\citenamefont{Huang, Chang, and
  Lin}}]{Huang2007}
\bibinfo{author}{\bibfnamefont{Y.~C.} \bibnamefont{Huang}},
  \bibinfo{author}{\bibfnamefont{C.~P.} \bibnamefont{Chang}}, \bibnamefont{and}
  \bibinfo{author}{\bibfnamefont{M.~F.} \bibnamefont{Lin}},
  \bibinfo{journal}{Nanotechn.} \textbf{\bibinfo{volume}{18}},
  \bibinfo{pages}{495401} (\bibinfo{year}{2007}).

\bibitem[{\citenamefont{Kumar et~al.}(2010)\citenamefont{Kumar, Jalil, Tan, and
  Liang}}]{Kumar2010}
\bibinfo{author}{\bibfnamefont{S.~B.} \bibnamefont{Kumar}},
  \bibinfo{author}{\bibfnamefont{M.~B.~A.} \bibnamefont{Jalil}},
  \bibinfo{author}{\bibfnamefont{S.~G.} \bibnamefont{Tan}}, \bibnamefont{and}
  \bibinfo{author}{\bibfnamefont{G.}~\bibnamefont{Liang}}, \bibinfo{journal}{J.
  Appl. Phys.} \textbf{\bibinfo{volume}{108}}, \bibinfo{pages}{033709}
  (\bibinfo{year}{2010}).

\bibitem[{\citenamefont{Petersen et~al.}(2011)\citenamefont{Petersen, Pedersen,
  and Jauho}}]{ACSnano2011}
\bibinfo{author}{\bibfnamefont{R.}~\bibnamefont{Petersen}},
  \bibinfo{author}{\bibfnamefont{T.~G.} \bibnamefont{Pedersen}},
  \bibnamefont{and} \bibinfo{author}{\bibfnamefont{A.-P.} \bibnamefont{Jauho}},
  \bibinfo{journal}{ACS Nano} \textbf{\bibinfo{volume}{5}},
  \bibinfo{pages}{523} (\bibinfo{year}{2011}).

\bibitem[{\citenamefont{Ouyang et~al.}(2011)\citenamefont{Ouyang, Peng, Liu,
  and Liu}}]{Ouyang2011}
\bibinfo{author}{\bibfnamefont{F.}~\bibnamefont{Ouyang}},
  \bibinfo{author}{\bibfnamefont{S.}~\bibnamefont{Peng}},
  \bibinfo{author}{\bibfnamefont{Z.}~\bibnamefont{Liu}}, \bibnamefont{and}
  \bibinfo{author}{\bibfnamefont{Z.}~\bibnamefont{Liu}}, \bibinfo{journal}{ACS
  Nano} \textbf{\bibinfo{volume}{5}}, \bibinfo{pages}{4023}
  (\bibinfo{year}{2011}).

\bibitem[{\citenamefont{MacKinnon}(1980)}]{MacKinnon1980}
\bibinfo{author}{\bibfnamefont{A.}~\bibnamefont{MacKinnon}},
  \bibinfo{journal}{J. Phys. C} \textbf{\bibinfo{volume}{13}},
  \bibinfo{pages}{L1031} (\bibinfo{year}{1980}).

\bibitem[{\citenamefont{MacKinnon}(1985)}]{MacKinnon1985}
\bibinfo{author}{\bibfnamefont{A.}~\bibnamefont{MacKinnon}},
  \bibinfo{journal}{Z. Phys. B} \textbf{\bibinfo{volume}{59}},
  \bibinfo{pages}{385} (\bibinfo{year}{1985}).

\bibitem[{\citenamefont{Novoselov et~al.}(2005)\citenamefont{Novoselov, Geim,
  Morozov, Jiang, Katsnelson, Grigorieva, Dubonos, and Firsov}}]{Novoselov2005}
\bibinfo{author}{\bibfnamefont{K.~S.} \bibnamefont{Novoselov}},
  \bibinfo{author}{\bibfnamefont{A.~K.} \bibnamefont{Geim}},
  \bibinfo{author}{\bibfnamefont{S.~V.} \bibnamefont{Morozov}},
  \bibinfo{author}{\bibfnamefont{D.}~\bibnamefont{Jiang}},
  \bibinfo{author}{\bibfnamefont{M.~I.} \bibnamefont{Katsnelson}},
  \bibinfo{author}{\bibfnamefont{I.~V.} \bibnamefont{Grigorieva}},
  \bibinfo{author}{\bibfnamefont{S.~V.} \bibnamefont{Dubonos}},
  \bibnamefont{and} \bibinfo{author}{\bibfnamefont{A.~A.}
  \bibnamefont{Firsov}}, \bibinfo{journal}{Nature (London)}
  \textbf{\bibinfo{volume}{438}}, \bibinfo{pages}{197} (\bibinfo{year}{2005}).

\bibitem[{\citenamefont{Zhang et~al.}(2005)\citenamefont{Zhang, Tan, Stormer,
  and Kim}}]{Zhang2005}
\bibinfo{author}{\bibfnamefont{Y.}~\bibnamefont{Zhang}},
  \bibinfo{author}{\bibfnamefont{Y.-W.} \bibnamefont{Tan}},
  \bibinfo{author}{\bibfnamefont{H.~L.} \bibnamefont{Stormer}},
  \bibnamefont{and} \bibinfo{author}{\bibfnamefont{P.}~\bibnamefont{Kim}},
  \bibinfo{journal}{Nature (London)} \textbf{\bibinfo{volume}{438}},
  \bibinfo{pages}{201} (\bibinfo{year}{2005}).

\bibitem[{\citenamefont{Gusynin and Sharapov}(2005)}]{Gusynin2005}
\bibinfo{author}{\bibfnamefont{V.~P.} \bibnamefont{Gusynin}} \bibnamefont{and}
  \bibinfo{author}{\bibfnamefont{S.~G.} \bibnamefont{Sharapov}},
  \bibinfo{journal}{Phys. Rev. Lett.} \textbf{\bibinfo{volume}{95}},
  \bibinfo{pages}{146801} (\bibinfo{year}{2005}).

\bibitem[{\citenamefont{Rhim and Park}(2012)}]{Rhim2012}
\bibinfo{author}{\bibfnamefont{J.-W.} \bibnamefont{Rhim}} \bibnamefont{and}
  \bibinfo{author}{\bibfnamefont{K.}~\bibnamefont{Park}},
  \bibinfo{journal}{Phys. Rev. B} \textbf{\bibinfo{volume}{86}},
  \bibinfo{pages}{235411} (\bibinfo{year}{2012}).

\bibitem[{\citenamefont{Gunst et~al.}(2011)\citenamefont{Gunst, Markussen,
  Jauho, and Brandbyge}}]{Gunst2011}
\bibinfo{author}{\bibfnamefont{T.}~\bibnamefont{Gunst}},
  \bibinfo{author}{\bibfnamefont{T.}~\bibnamefont{Markussen}},
  \bibinfo{author}{\bibfnamefont{A.-P.} \bibnamefont{Jauho}}, \bibnamefont{and}
  \bibinfo{author}{\bibfnamefont{M.}~\bibnamefont{Brandbyge}},
  \bibinfo{journal}{Phys. Rev. B} \textbf{\bibinfo{volume}{84}},
  \bibinfo{pages}{155449} (\bibinfo{year}{2011}).

\bibitem[{\citenamefont{Zhang et~al.}(2013)\citenamefont{Zhang, Lu, Shi, Wang,
  Zhang, Sun, Zheng, Chen, Wang, Lin et~al.}}]{Zhang2013}
\bibinfo{author}{\bibfnamefont{H.}~\bibnamefont{Zhang}},
  \bibinfo{author}{\bibfnamefont{J.}~\bibnamefont{Lu}},
  \bibinfo{author}{\bibfnamefont{W.}~\bibnamefont{Shi}},
  \bibinfo{author}{\bibfnamefont{Z.}~\bibnamefont{Wang}},
  \bibinfo{author}{\bibfnamefont{T.}~\bibnamefont{Zhang}},
  \bibinfo{author}{\bibfnamefont{M.}~\bibnamefont{Sun}},
  \bibinfo{author}{\bibfnamefont{Y.}~\bibnamefont{Zheng}},
  \bibinfo{author}{\bibfnamefont{Q.}~\bibnamefont{Chen}},
  \bibinfo{author}{\bibfnamefont{N.}~\bibnamefont{Wang}},
  \bibinfo{author}{\bibfnamefont{J.-J.} \bibnamefont{Lin}},
  \bibnamefont{et~al.}, \bibinfo{journal}{Phys. Rev. Lett.}
  \textbf{\bibinfo{volume}{110}}, \bibinfo{pages}{066805}
  (\bibinfo{year}{2013}).

\end{thebibliography}
\end{document}